\documentclass[12pt]{iopart}

\usepackage{iopams}
\usepackage{color}
\usepackage{graphicx}
\usepackage{units}

\begin{document}

\title[SAW mediated carrier injection into remote self-assembled QPs]{Surface acoustic wave mediated carrier injection into individual quantum post nano emitters}
\author{Stefan V\"olk$^{1,2,\dag}$, Florian Knall$^{1,2,\dag}$, Florian J. R. Sch\"ulein$^{1,2}$, Tuan A. Truong$^3$, Hyochul Kim$^{3,4}$, Pierre M. Petroff${^3}$, Achim Wixforth$^{1,2}$, Hubert J. Krenner$^{1,2,*}$}
\address{$^1$ Lehrstuhl f\"ur Experimentalphysik 1 and Augsburg Centre for Innovative Technologies {\it ACIT}, Universit\"at Augsburg, 86159 Augsburg}
\address{$^2$ Center for NanoScience {\it CeNS}, Geschwister-Scholl-Platz 1, 80539 M\"{u}nchen, Germany}
\address{$^3$ Materials Department, University of California, Santa Barbara CA, 93106, United States}
\address{$^4$ Department of Electrical and Computer Engineering, IREAP, University of Maryland, College Park, MD 20742, United States}
\address{$^\dag$ These authors contributed equally to this work.}
\ead{$^*$ hubert.krenner@physik.uni-augsburg.de}

\begin{abstract}
Acousto-electric charge conveyance induced by a surface acoustic wave is employed to dissociate photogenerated excitons. Over macroscopic distances, both electrons and holes are injected  \emph{sequentially} into a remotely positioned, isolated and high-quality quantum emitter, a self-assembled quantum post. This process is found to be highly efficient and to exhibit improved stability at high acoustic powers when compared to direct optical pumping at the position of the quantum post. These characteristics are attributed to the wide matrix quantum well in which charge conveyance occurs and to the larger number of carriers available for injection in the remote configuration, respectively. The emission of such pumped quantum posts is dominated by recombination of neutral excitons and fully directional when the propagation direction of the SAW and the position of the quantum post are reversed. 
\end{abstract}


\maketitle

Ambipolar acousto-electric charge conveyance of dissociated electrons and holes in the plane of a quantum well (QW) based on piezoelectric surface acoustic waves (SAWs) \cite{Rocke:97} has been proposed to realize a precisely triggered single photon source operating at radio frequencies in the range from a few tens of megahertz up to several gigahertz. In this scheme, electrons and holes are transported in the type-II potential modulation induced by the SAW and sequentially injected into a quantum dot (QD) which emits a single photon per SAW cycle \cite{Wiele:98}. Until now, this remote injection scheme has been investigated using QD systems of moderate optical quality or weak confinement potential such as QDs induced in surface-near QWs by stressor islands on the sample surface\cite{Sopanen:1995,Schuelein:09,Boedefeld:06}, interface fluctuation QDs in disordered quantum wires (QWRs) \cite{Notzel:98,Couto:09} or Indium-rich segments in GaAs nanowires (NWs) \cite{Hernandez:12}. Photon correlation spectroscopy performed on sharp emission lines of localized QD-like emission centers in acoustically pumped QWRs and NWs revealed signatures of \emph{multiple} emission centers contributing to the detected signals. These limitations arise from the weak confinement potential of these disorder based approaches \cite{Lazic2012}. Thus, alternative routes based on established self-assembled quantum dot nanostructures are required for which clear photon anti-bunching has been routinely observed \cite{Michler:00,Shields:07}. Moreover, these QDs can be embedded in high quality factor optical nanocavities to fully exploit the high gigahertz repetition frequencies provided by SAWs using the Purcell enhancement of the spontaneous emission rate \cite{Strauf:07}. In conventional self-assembled QDs, the confined electronic states and the resulting emission energies of excitonic complexes can be efficiently manipulated by a SAW \cite{Gell:08,Metcalfe:10}. For the implementation of the envisioned remote acousto-electric injection scheme, a high quality two-dimensional QW is crucially required to prevent carrier losses in the vertical direction.  The typical two-dimensional wetting layer on which these islands nucleate indeed provides a QW-like confinement potential, however, due to its small thickness of $<1$\,nm, acousto-electric charge conveyance is inhibited by the resulting low carrier mobilities \cite{Voelk:10b}. One alternative system are height controlled, self-assembled quantum posts (QPs) \cite{He:07}, the height of which, in contrast to  their QD counterparts, can be controlled from $\sim3$\,nm up to $>50$\,nm while preserving high quality single photon emission \cite{Krenner:08a,Krenner:09}. Here, we report the direct observation of SAW-mediated electron-hole injection into isolated, individual self-assembled QPs. Since these QPs are embedded in a wide lateral QW, the acousto-electric charge conveyance by the SAW is highly efficient and remote carrier injection is demonstrated over macroscopic distances only limited by the optically accessible area. After optical generation, electron-hole pairs are dissociated by the SAW and transported to the QP. The two carrier species arrive with a fixed time delay of $\Delta t=2.6$\,ns. This unique approach guarantees an \emph{inherently sequential injection} of electrons and holes, a property which is challenging to achieve e.g. using Coulomb blockade. This mechanism is found to be highly directional with improved stability even at highest SAW amplitudes in contrast to a configuration where carriers are generated at the position of the emitter. Our results represent a major step towards a precisely-triggered SAW-driven single photon source using high quality quantum emitters.\\

The sample studied was grown by solid-source molecular beam epitaxy (MBE) on a semi-insulating (001)-GaAs substrate. A single layer of $\sim23$\,nm high QPs was realized using eight repetitions of a growth cycle consisting of depositions of 1 monolayer of InAs and 7 monolayers of GaAs, each followed by 1 minute growth interruptions \cite{He:07}. The sample was finalized by a 100\,nm GaAs capping layer. During QP growth, the substrate rotation was interrupted giving rise to a coverage gradient and a variation of the QP surface density on the substrate \cite{Krenner:05a}. In the region where island nucleation breaks down, ultra-low QP surface densities of $<0.005$ QPs per $\mu$m$^2$ are realized. Such ultra-low surface densities allow for optical investigation of individual QPs under both direct optical excitation and under remote SAW-driven carrier injection since the average QP separation of $\gtrsim15$\,$\mu$m is comparable to or larger than typical SAW wavelengths. On these samples we fabricated interdigital transducer (IDT) electrodes for excitation of SAWs of wavelength $\lambda_{\rm SAW}=15$\,$\mu$m, corresponding to SAW frequencies $f_{\rm SAW}=193$\,MHz at low temperatures $(T=6$\,K$)$. We studied these samples by conventional low-temperature micro-photoluminescence spectroscopy. For optical excitation we used a diode laser emitting $\lesssim 100$\,ps pulses $(f_{\rm laser}=80$\,MHz) with a photon energy of $E_{\rm laser}=1.875$\,eV focused by a $50\times$ microscope objective to a $\sim2$\,$\mu$m spot to photogenerate electron-hole pairs in the GaAs continuum. The emitted luminescence of the sample was collected over the full field of view of the objective and sent to a 0.5\,m imaging monochromator equipped with a liquid nitrogen cooled two-dimensional CCD detector for spectral and spatial analysis. Due to the time-integrated detection and the unlocked excitation scheme $(f_{\rm SAW}\sim 2.4f_{\rm laser})$, the recorded data provides an averaged picture of the SAW-controlled carrier injection and recombination. For all experiments we employed low optical pump powers for which screening of the SAW-induced potential by the photogenerated carriers can be neglected.\\

\begin{figure}[ht]
	\begin{center}
		\includegraphics[width=0.8\textwidth]{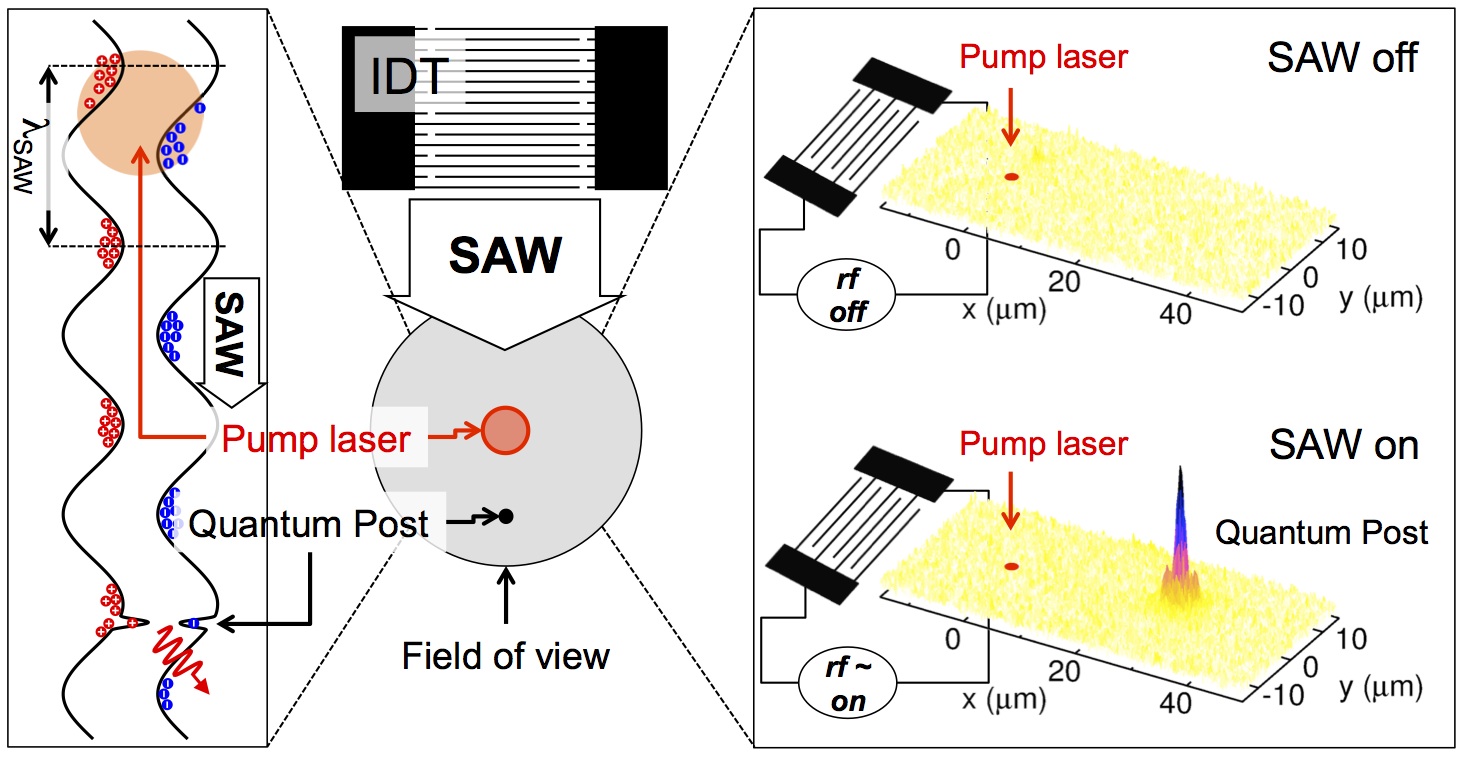}
		\caption{-- {\bf Remote acousto-electric carrier injection} -- Center: An IDT generates a SAW which propagates through the field of view of the objective lens (gray-shaded area) with the pump laser focused in its center and a single QP shifted along the SAW propagation direction. Left: Type-II band edge modulation induced by the SAW leads to a spatial separation of electrons and holes and after acousto-electric charge conveyance to \emph{sequential injection} into the QP. Right: PL images of the QP emission  without and with a SAW generated. The pump laser spot is located in the center. The QP emission isolated by a 900 nm longpass interference filter. A diffraction limited emission spot is observed with the SAW applied due to acousto-electric charge transport over a distance of 30\,$\rm \mu m$ and subsequent injection into a single QP.}
		\label{fig:1}
	\end{center}
\end{figure}

The acousto-electric charge injection scheme is depicted schematically in Fig. \ref{fig:1}. In the center of the field of view (gray-shaded area), a focused laser (red circle) generates electron-hole pairs. These are dissociated in the type-II modulation of the effective band edges of the QW by the SAW \cite{Rocke:97} as depicted schematically on the lefthand side of Fig. \ref{fig:1}. The SAW itself is electrically generated by applying an rf voltage to an IDT and transports electrons and holes along its propagation direction. A single QP acts as a deep confinement potential in the conduction and valence bands into which electrons and holes are sequentially injected \cite{Wiele:98,Boedefeld:06,Couto:09}. An exciton is formed after each full cycle of the SAW which radiatively recombines by emission of a single photon. On the righthand side of Fig. \ref{fig:1} we present direct (spatial) images of the sample emission without and with a SAW applied. The excitation laser $(P_{\rm laser}= 72$\,nW$)$ was focused at the center of the two images at $x=y=0$ as marked by a red circle. The detected intensity is encoded in false color with yellow (black) corresponding to low (high) signal levels. Here we used a 900\,nm long pass to ensure detection only of emission originating from exciton recombination inside QPs. Clearly, with the SAW turned off, no signal is detected over the entire area. When a SAW is excited at $P_{\rm rf}=+18$\,dBm on the lefthand side as indicated schematically by the IDT (not to scale), we observe a bright, diffraction limited, point-like emission spot arising from emission of a single, isolated QP located at $(x=30$\,$\rm \mu m$, $y=2$\,$\rm \mu m)$. This experiment directly confirms directional SAW-mediated acousto-electric charge conveyance over a distance of two acoustic wavelengths and subsequent injection into an isolated zero-dimensional quantum emitter. 

\begin{figure}[ht]
	\begin{center}
		\includegraphics[width=0.8\textwidth]{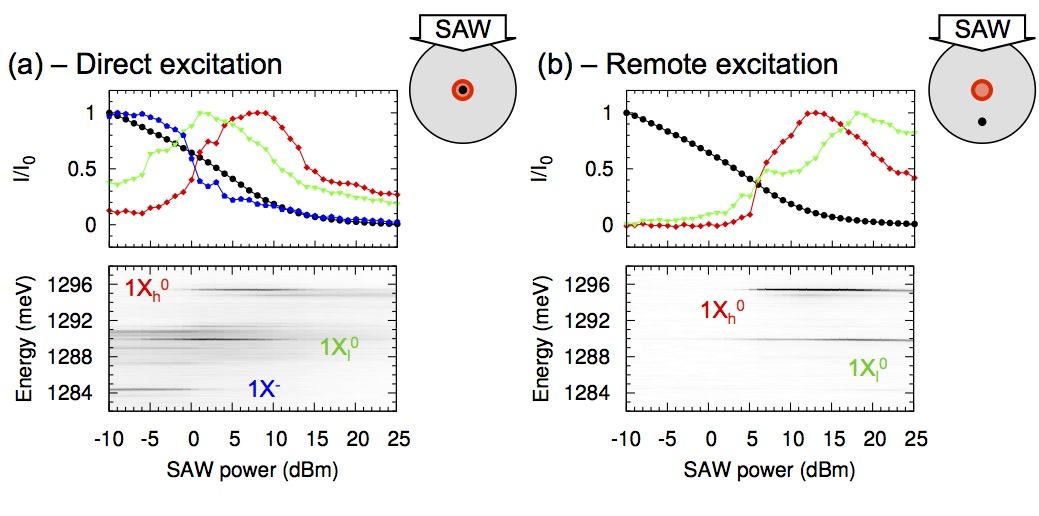}
		\caption{-- {\bf SAW power dependent emission under direct and remote charier injection} -- Gray-scale plots of PL intensity (lower panels) as a function of SAW power under direct (a) and remote injection (b) as indicated by the schematics with the corresponding exciton configuration labeled. The normalized intensities of $1X^0_{l}$ (green triangles/line), $1X^0_{h}$ (red squares/line) and $1X^-$ (blue pentagons/lines)  are plotted in the upper panels. The quenching of QW emission (black bullets/line) is accompanied by a change-over to neutral exciton emission under direct excitation (a) and remote carrier injection (b).}
		\label{fig:2}
	\end{center}
\end{figure}

In the following, we investigate SAW-controlled carrier injection in detail. We distinguish between two different configurations, \emph{direct injection} for which the excitation laser is focused at the position of the QPs and \emph{remote, time delayed and sequential} injection for which the point of carrier generation and the QP are spatially separated. In Fig. \ref{fig:2} (a) and (b) we present emission spectra under weak optical pumping $(P_{\rm laser}= 22$\,nW$)$ of the same, single QP under direct and remote carrier injection, respectively, as indicated by the associated schematics. The distance under remote injection was $d=-20$\,$\rm \mu m$, corresponding to more than 1.3 acoustic wavelengths, ensuring carrier injection occurs only due to acousto-electric charge conveyance. The lower panels show the detected QP emission encoded in grayscale (white low, black high signal) as a function of the rf power applied to the IDT. For the directly excited QP [cf. Fig. \ref{fig:2} (a)] we observe the characteristic intensity change-over or switching behavior of dominant emission from the negatively charged exciton $(1X^-=2e+1h)$ to charge neutral single excitons $(1X^0_{l,h}=1e+1h)$ \cite{Voelk:10b,Voelk:11a}. For QPs, we mostly observe doublets with indices $l$ and $h$ denoting the lower and higher energy line. These arise from localized hole states at the two ends of the QP which is confirmed by the observation of avoided crossing in electric field tunable structures \cite{Krenner:08b} and selective injection under SAW pumping \cite{Voelk:11a}. The absolute energy spacing of this doublet depends on the actual morphology of the QPs and is comparably large for the example presented here. The switching occurs in the rf power range between 0\,dBm and +5\,dBm as illustrated by the extracted integrated and normalized intensities in the upper panel. The $1X^-$ emission (blue pentagons/lines) is largely suppressed at $P_{\rm rf}=0$\,dBm and is replaced by emission of the two $1X^0$ emission lines (red squares/line and green triangles/line). The onset of acousto-electric charge conveyance is further confirmed by the quenching of the QW PL in the same power range which is plotted as a reference (black bullets/line) in the upper panel of Fig. \ref{fig:2} (a) and (b). For high rf power levels, $P_{\rm rf}>+12$\,dBm, we observe a pronounced quenching due to increasingly more efficient acousto-electric transport of the photo-excited carriers away from the QP position. 
\begin{figure}[ht]
	\begin{center}
		\includegraphics[width=0.5\textwidth]{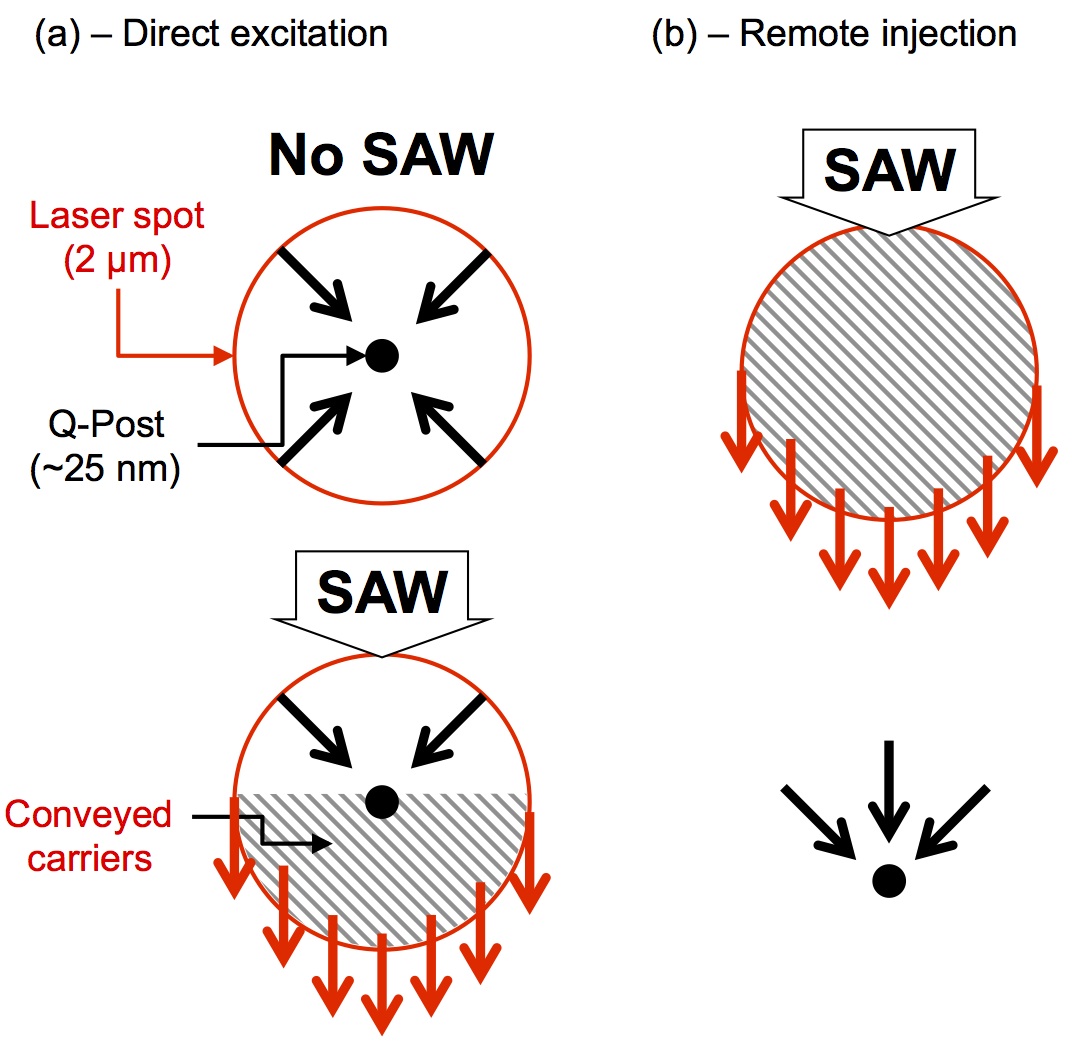}
		\caption{-- {\bf Fractions of carriers available for injection into QP} -- (a) For direct excitation the number of carriers available for injection (black arrows) into the QP is reduced compared to no SAW applied due to acousto-electric charge conveyance (red arrows) of carriers generated in the hatched area. (b) For remote excitation electrons and holes generated in the entire area of the pump laser can be transported (red arrows) and can be injected (black arrows) into the QP leading to a reduced quenching of the emission at high acoustic powers.}
		\label{fig:3}
	\end{center}
\end{figure}
This is depicted schematically in the upper panel of Fig. \ref{fig:3} (a), where at low acoustic powers carriers being photogenerated within the $2$\,$\rm \mu m$ laser spot can be captured into the QP as indicated by the black arrows. In contrast, at the highest acoustic power levels applied, a significant fraction photogenerated charges are transported out of the generation area by the SAW indicated by the red arrows. Under these conditions approximately half the photogenerated carriers created, predominantly from the hatched area in the lower panel Fig. \ref{fig:3} (a), are transported away from the position of the QP and cannot be captured. This net decrease of the number of available carriers gives rise to the observed pronounced reduction of the observed total emission intensity.\\

For the case of remote acoustic carrier injection [cf. Fig. \ref{fig:2} (b)] into a QP located $20$\,$\rm \mu m$ away from the position of the laser focus we observe no emission from the QP up to $P_{\rm rf}=0$\,dBm. As acoustic charge conveyance sets in and carriers are transported by the SAW to the QP, we detect emission arising from recombination of \emph{only} $1X^0_{l}$ and  $1X^0_{h}$ states which gradually increases up the $P_{\rm rf}=+12$\,dBm. Since $1X^-$ generation is efficiently suppressed under sequential electron-hole injection, the corresponding emission is absent in the data of the remotely pumped QP. At high acoustic power levels, we find two pronounced differences compared to the directly excited QP shown in Fig. \ref{fig:2} (a): (i) The lower energetic branch of $1X^0_{l}$ becomes the dominant emission line under remote carrier excitation and (ii) the overall decrease of intensity is significantly reduced compared to the directly excited QP. 
The first observation can be readily understood by taking into account that holes are transported in a maximum of the type-II band edge modulation [cf. Fig. \ref{fig:1}]. At this phase of a Rayleigh type SAW, the lateral electric field component is zero while the vertical component is maximum and oriented upwards along the growth direction. Thus, holes are conveyed preferentially on the surface-near side of the matrix QW and preferentially injected in the upper end of the QP \cite{Voelk:11a}. This effect is expected to become more pronounced at high SAW amplitudes and under strict sequential injection. Therefore, this vertical electric field effect is best resolved under remote injection even at the relatively small separation between the pump laser spot and the QP of $1.2\,lambda_{\rm SAW}$. The dissimilar quenching behavior cannot be explained by tunnel escape of electrons which is expected to be weak for typical SAW fields of $<10$\,kV/cm \cite{Fry:00b,Kaniber:11} and should be the same under both pumping conditions. The weaker quenching behavior can be understood qualitatively by taking into account, that under remote injection the QP is located outside the generation area as shown in Fig. \ref{fig:3} (b). Thus, net a larger number of photogenerated carriers is transported to the QP position and can be captured in its quantized energy levels in strong contrast to the case of direct excitation. In this simple picture, lateral spreading of carriers perpendicular to the SAW propagation direction is not taken into account. This process, which so far has not been included in numerical simulations of acousto-electric charge conveyance \cite{Garcia:04}, was found to be fast for electrons \cite{Krauss:02} and could be in parts responsible for the suppression of the $1X^-$ emission \cite{Voelk:10b}. However, since this effect decreases the electron density at the position of the QP, we can exclude it as a \emph{dominant} mechanism for the geometry studied. To fully understand the underlying microscopic processes, detailed studies of the remote carrier injection with time-resolved detection schemes would be required. Such experiments performed as a function of the pump laser-QP separation and the applied rf power could then be compared to numerical simulations in a two-dimensional extension of the semi-classical transport model introduced by Garc\'{i}a-Crist\'{o}bal et al. \cite{Garcia:04}. \\

\begin{figure}[ht]
	\begin{center}
		\includegraphics[width=0.8\textwidth]{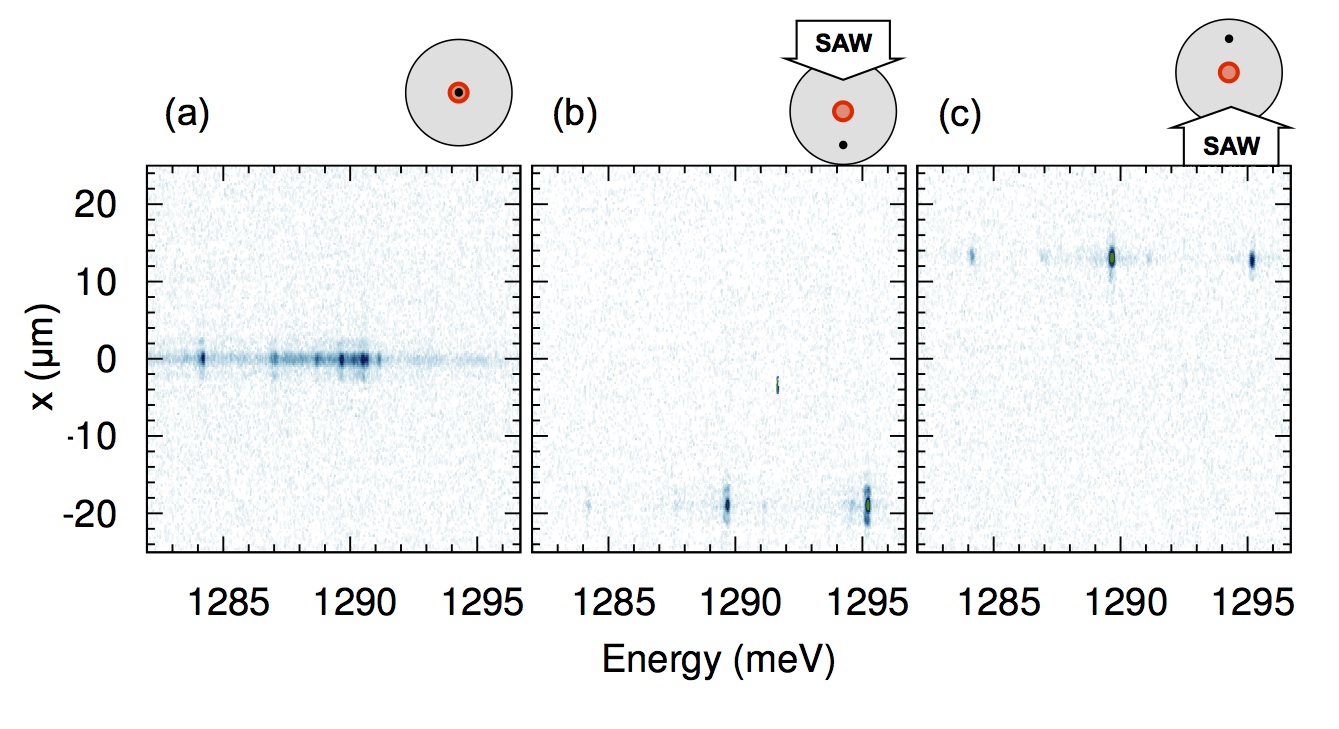}
		\caption{-- {\bf Directional SAW driven carrier injection} -- Spatially resolved PL spectra of the same QP under direct optical excitation with no SAW applied (a) and remote injection with SAW propagating downwards (b) or upwards (c) and the QP spatially translated into the respective direction as indicated by the schematics.}
		\label{fig:4}
	\end{center}
\end{figure}

Finally, we address the directionality of the remote acousto-electric charge conveyance. In order to obtain \emph{spatial and spectral} information, we do not integrate over the entire CCD array but read out the spatial information encoded in the rows of the array while the spectral information is contained in the columns of the array. We mounted the sample such that the SAW propagation direction ($P_{\rm rf}=+10$\,dBm) was aligned with the CCD columns. In Fig. \ref{fig:4} (a), we show a false color PL image recorded from the \emph{same} QP studied previously under direct optical excitation $(P_{\rm laser}= 22$\,nW$)$ and without a SAW applied. Clearly, emission is only found in the center row of the array marking the reference position of the laser focus. In a first step, we keep the optical pumping power constant and moved the QP downwards. With a SAW propagating from top to bottom [Fig. \ref{fig:4} (b)], acoustic charge conveyance transports the photogenerated electrons and holes towards the remotely positioned QP. Due to remote sequential carrier injection we detect the QP emission $d=-20$\,$\rm \mu m$ below the point of optical charge generation in the corresponding PL image. All data shown in Fig. \ref{fig:2} (b) was recorded in this configuration. Moreover, the switching from $1X^-$ to $1X^0_l$ and $1X^0_h$ discussed in the previous section is clearly resolved in the energy-domain. In a second step, moving the QP upwards and exciting the SAW propagating from bottom to top using the same acoustic power, the spatial configuration is reversed [cf. schematic with Fig. \ref{fig:4} (c)]. In the experimental data shown in Fig. \ref{fig:4} (c), we observe the same emission pattern, characteristic for sequential electron/hole injection, but translated in space to $d=+15$\,$\rm \mu m$. Close examination of the relative intensities reveal a slight modification of the relative intensities of $1X^0_l$ and $1X^0_h$ which could arise from the smaller separation of $\sim1\lambda_{\rm SAW}$ between the QP and the pump laser spot in this configuration \cite{Hernandez:12}. This confirms acousto-electric charge conveyance from the center of the image upwards to the position of the QP. Thus, our SAW-based injection scheme allows for selective addressing of individual QPs using different SAW propagation directions. In addition, multiple QPs with distinct and characteristic emission energies can be pumped by a single SAW beam when located along the same propagation direction. To further exploit the directionality of this mechanism, NW-based approaches \cite{Hernandez:12} could be extended to single crystal phase structures \cite{Kinzel:11} with embedded axial QDs \cite{Tribu:08a,Kouwen:10a}.\\

In conclusion, we demonstrated directional, remote carrier injection into isolated self-assembled QPs employing acousto-electric charge conveyance induced by a SAW. Our findings demonstrate that QP with their surrounding wide matrix QW are an ideal system for the realization of a precisely triggered SAW-controlled single photon source \cite{Wiele:98}. In order to fully exploit the potential of this unique approach, further steps have to be taken, namely the combination with high quality optical nanocavities to achieve Purcell-enhancement of the radiative emission rate and improved photon extraction. The latter is currently the limiting factor in order to directly observe acoustically pumped single photon emission which has been clearly observed before for direct injection \cite{Krenner:08a}. QPs have been readily coupled to DBR-microcavity lasers \cite{Kim:09} which are fully compatible with SAW techniques \cite{Lima:05} and also more advanced approaches based on oxide aperture \cite{Strauf:07} or photonic crystal based structures \cite{Fuhrmann:11} are now within reach. For all-electrical operation of this type of single photon source lateral {\it p-i-n}-junctions \cite{Stavarache:06} will be used to inject electrons and holes from opposite sides into the propagation path of the SAW. An even more tantalizing direction is the extension to high-fidelity entangled photon pair generation using the biexciton-exciton cascade. Since charged exciton formation is largely suppressed under direct \cite{Voelk:10b} and remote sequential electron-hole injection, our SAW-based approach could circumvent challenges arising in the originally envisioned Coulomb-blockade controlled carrier injection \cite{Benson:00} scheme. For this approach self-assembled GaAs/AlGaAs QDs grown inside etched nanoholes \cite{Rastelli:04a,Wang:09a} might be superior. For such QDs, the thickness of the wetting layer can be arbitrarily increased for optimal acousto-electric charge transport, no emission doublets due to hole localization are observed. Moreover, the emission wavelengths of such QDs are blue-shifted to regions of higher sensitivity of Si-based detectors. 

\section*{Acknowledgements}
This work was supported by the cluster of excellence "Nanosystems Initiative Munich" (NIM), by DFG via the Emmy Noether Program (KR 3790/2-1) and Sonderforschungsbereich 631 and by the Alexander-von-Humboldt Foundation.

\section*{References}
\bibliographystyle{iopart-num}
\bibliography{library}

\end{document}